\documentclass[%
 reprint,
 amsmath,amssymb,
 aps,
]{revtex4-2}
\usepackage{amsmath}
\usepackage[toc]{appendix}
\usepackage{graphicx}
\usepackage{dcolumn}
\usepackage{comment}
\usepackage{bm}
\usepackage{float}
\usepackage[dvipsnames]{xcolor}
\usepackage{mathcomp}

\usepackage[hidelinks]{hyperref}

\newcommand{\MR}[1]{{\color{Orange}{[MR: #1]}}}

\usepackage{booktabs}

\DeclareUnicodeCharacter{2212}{-}

\newcommand\scalemath[2]{\scalebox{#1}{\mbox{\ensuremath{\displaystyle #2}}}}

\newcommand{\spnote}[1]{#1}

\begin{document}
\preprint{APS/123-QED}

\author{Ruben Gargiulo}%
 \email{ruben.gargiulo@uniroma1.it}
\affiliation{Università degli Studi La Sapienza, Piazzale Aldo Moro 5, 00185 Roma, Italy}%

\author{Elisa Di Meco}
 \affiliation{INFN Laboratori Nazionali di Frascati, Via Enrico Fermi 54, 00044 Frascati, Italy}

\author{Stefano Palmisano}
\affiliation{Università degli Studi La Sapienza, Piazzale Aldo Moro 5, 00185 Roma, Italy}

\date{\today}
             
\title{Feasibility study of True Muonium discovery with CERN-SPS H4 positron beam}




\date{\today}

\begin{abstract}
True muonium ($\mu^+\mu^-$) is one of the heaviest and smallest electromagnetic bound states not containing hadrons, and has never been observed so far. 
\spnote{In this work we }show that the spin-1 TM  state (ortho-TM) can be observed at a discovery level of significance in three months at the CERN SPS North-Area H4A beam line, using 43.7 GeV secondary positrons. In this way, by impinging the positrons on multiple thin low-Z targets, ortho-TM, which decays predominantly to $e^+e^-$, can be produced from $e^+e^- \to TM$ interaction on resonance ($\sqrt{s} \sim 2m_{\mu}$).

\end{abstract}

\maketitle


\section{\label{sec:intro} Introduction}

Quantum electrodynamics (QED) predicts the existence of several bound states, in addition to standard atoms, such as purely leptonic systems.
The lightest one, the positronium ($e^+e^-$), has been discovered decades ago and extensively studied \cite{positronio}.

In contrast, true muonium ($\mu^+\mu^-$) and true tauonium ($\tau^+\tau^-$) have never been observed, due to the lack of $e^+e$- colliders running at the proper center-of-mass energy to exploit the enhanced resonant cross-section. In addition, dissociation effects in matter complicate detection in fixed-target experiments.
Focusing on true muonium (TM), there are several known pathways for its discovery.
The simplest one, from a theoretical point of view, is the resonant production of orto-TM (spin 1 state, decaying into $e^+e^-$) in $e^+e^-$ collisions.
Its 66.6 nb peak cross-section allows the observation of \(TM \to e^+e^-\) through displaced decay vertices \cite{Brodsky:2009gx}:
    \begin{itemize}
        \item at new dedicated colliders:
        \begin{itemize}
            \item using large-angle collisions and $O(1)$ GeV beam energies, creating boosted and easily observable TM atoms, but facing difficulties in building a dedicated collider with a particular geometry \cite{dimus} \cite{russi}
            \item using normal small-angle collisions at the proper $\sqrt{s} \approx 2m_{\mu}$ center-of-mass energy and observing only the TM excited states with longer lifetimes \cite{gargiulo2024true}
        \end{itemize}
        \item at fixed-target experiments with positron beams of $\sim 43.7$ GeV (readily available at the CERN North-Area H4A beam line facility), to have $e^+e^-$ collisions at $\sqrt{s} \approx 2m_{\mu}$ \cite{crivelli}
    \end{itemize}
    
    Another option with $e^+e^-$ interactions is out-of-resonance radiative $e^+e^- \to TM \gamma$ production \cite{Brodsky:2009gx} at existing colliders running at $O(1)$ GeV center-of-mass energies, with very low $O(10^{-1})$ fb cross-sections.
    
    It is also possible to create TM from mesons decays \cite{redtop, malenfant}:
    \begin{itemize}
        \item $\eta \to TM \gamma$, with a branching ratio of $\sim 5 \times 10^{-10}$ \cite{lamm-ji}. This production mode will be searched for at LHCb starting from the large $pp \to \eta X$ sample \cite{lhcb}
        \item $K_{L} \to TM \gamma$, with a branching ratio of $\sim 7 \times 10^{-13}$ \cite{PhysRevD.98.053008}, searchable by future neutral-kaon based high-intensity experiments 
    \end{itemize}
    Other possibilities are:
    \begin{itemize}
    \item Bremmstrhalung-like and triplet-like processes in electron-nucleus scattering ($eZ \to e \,TM\,Z$) \cite{Holvik:1986ty}, with extremely low $O(10^{-2})$ fb cross-section \cite{xfel}
    \item Photon-photon fusion in relativistic heavy ion collisions \cite{Ginzburg:1998df}, with reasonably high $O(1)$ $\mu$b cross-section. Note that for heavy ion colliders (such as LHC) the luminosity ranges in the $O(1)$ nb$^{-1}$ per year region.
    \item Interactions of ultra-slow negative and positive muons within a target, with interactions of $\mu^-$ with muonium ($\mu^+ e^-$) or $\mu^+$ with muonic hydrogen ($\mu^- p$) \cite{sakimoto}
\end{itemize}

Among all these methods, the only ones having a potentially fast timescale are fixed-target resonant $e^+e^-$ production and $\eta$ meson decay based production. As a matter of fact, they do not require the construction of new colliders or beam facilities.
TM production from $\eta$ meson decay only requires the LHCb Run 3 data to be analysed starting in 2025, while fixed-target resonant $e^+e^-\to TM$ production can be realized in three months of data-taking in a readily available experimental hall (H4 beam line at the CERN North Area) with standard detector technologies using a 43.7 GeV positron beam.

This paper focuses on this last option and includes theoretical calculations of the effective event yield, discussions on target geometry, beam and detector requirements, and the discovery potential.

\section{True Muonium properties}
\begin{figure}[H]
\centering
    \includegraphics[width=\columnwidth]{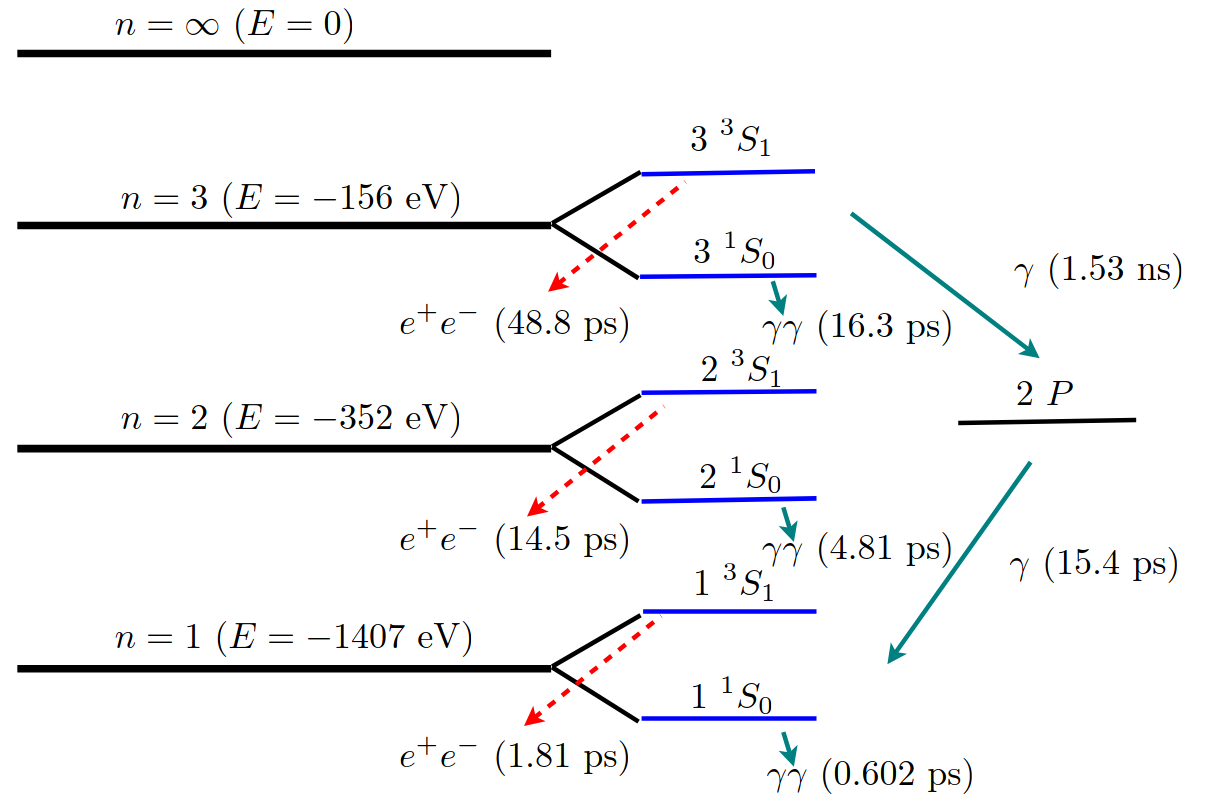}
    \caption{True muonium levels, lifetimes and transitions diagram for $n\leq 3$ (spacing not to scale)~\cite{Brodsky:2009gx}.}
    \label{fig:TM}
\end{figure}

TM energy levels can be calculated by rescaling the positronium spectrum: the binding energy of the deepest level (1S) is $\mathrm{B.E.(1S)} = 1.4$ keV, as shown in Figure \ref{fig:TM}.\\
Like positronium, TM has two spin states: para-TM (spin 0), which decays to $\gamma \gamma$, and orto-TM (spin 1), which decays \spnote{predominantly} to $e^+e^-$ \cite{Brodsky:2009gx}.\\
The lifetimes of the n-th $S$ levels for the two spin states $s=0, \, 1$ are proportional to $n^3$ (at lowest order), as follows \cite{Brodsky:2009gx}:
\begin{align}
    \tau(nS_{s=1} \to e^+e^-) = \frac{6 \hbar n^3}{\alpha^5 m_\mu c^2} \sim n^3 \times 1.8 \, \mathrm{ps} \\
    \tau(nS_{s=0} \to \gamma \gamma) = \frac{1}{3} \tau(nS_{s=1} \to e^+e^-)\,.
\end{align}
%
These lifetimes are much shorter than the muon lifetime; therefore, the muons in the TM can be considered stable particles.

\section{$e^+e^- \to TM$ production cross-section at fixed-target}


The total production cross-section for true muonium on resonance in $e^+e^-$ scattering reads \cite{Brodsky:2009gx}:

\begin{equation}\label{res}
    \sigma_{ON\,R.} = 2\pi^2 \frac{\alpha^3}{s} = \frac{ \pi^2 \alpha^3}{2 m_{\mu}^2} = 66.6 \, \mathrm{nb}\,.
\end{equation}

The probability to produce TM in a state $n$ is proportional to $n^{-3}$ \cite{Brodsky:2009gx}, and the normalization factor is $\zeta(3)$, where:
\begin{equation}
    \zeta(k) = \sum_{n=1}^{+\infty} \frac{1}{n^k}
\end{equation}
is the Riemann Zeta function.

Hence, the probability to produce the TM in the ground state 1S is $\epsilon_{1S} = 83\%$.

In order to address TM production in real conditions, with non-negligible fluctuations in the $\sqrt{s}$, the cross-section in Eq. \eqref{res} need to be reduced accounting for the probability $p$ \spnote{($p$-factor)} that the beam center-of-mass energy is in the range $(2m_\mu - B.E.(1S), \, 2m_\mu)$ where bound states are allowed \cite{Brodsky:2009gx}. 
To obtain a precise value of the cross-section, the effect of initial state radiation must also be included.


\subsection{Matter effects}
With respect to colliders, fixed-target resonant production presents two additional effects due to the presence of a solid target.

The dominant process in the interaction of TM with target material is dissociation into $\mu^+\mu^-$ pair. The dissociation cross-section is very large ($\sigma_d \sim 13 Z^2$\,b) as pointed out in \cite{lhcb}.


Incoherent interactions with matter can also cause TM to change its spin, from the ortho form, produced in the $e^+e^-$ collision, to the para form.
The subsequent decay of the para-TM into $\gamma \gamma$, instead of the expected $e^+e^-$, can cause detection inefficiencies \cite{relatoms}.
The ortho-para transitions have a cross-sections $O(1)$ mb, a factor at least $10^4$ smaller than the dissociation cross-section. For this reason this effect can be safely neglected.

The expected TM yield per impinging positron for a target of thickness $\Delta z$, and $\epsilon$ global detection efficiency\spnote{, including the $p$-factor (see next sections)} is:
\begin{align}
    \frac{dTM}{de^+ dN_{\mathrm{target}} } = \epsilon N_A \rho Z p \sigma_{ON R.} \int_{0}^{\Delta z} dz e^{-\mu_d z} = \\
    = \epsilon \frac{p \, \sigma_{ON R.}}{13 Z \, \mathrm{b}} (1 - e^{-\Delta z \mu_d})
    \label{ntm}
\end{align}
where $\rho$ is the atomic density and $\mu_{d} = N_A \rho \sigma_{d}$ is the inverse dissociation length.\\
The formula underlines two important facts:
Firstly, the event yield as a function of the target thickness saturates for large thickness values, reaching an upper limit depending only on the material. 
Secondly, low-Z targets produce higher yield.
The natural material choice for the target is therefore lithium, which is the lightest element available in metallic foils. Lithium is highly reactive, so it should be kept under inert gas, mineral oil, or vacuum. This requirement is not particularly problematic from a technological standpoint. 

Going back to interactions of TM with matter, the second very important effect is the non-negligible fluctuation of the momenta of the target electrons, due to atomic bonds.
As recently pointed out in \cite{Arias-Aragon:2024qji} the electron motion in the target material can affect the effective production cross-section significantly by changing the actual $\sqrt{s}$ of the collision with respect to the electron at rest hypotheses.

Taking $k_{-},\,E_{-},\,\beta_{-},\,\gamma_{-}$ as the electron kinematical variables, and $\theta_{-}$ as the electron angle with respect to the beam direction during the collision, the resulting center-of-mass energy is:
\begin{equation}
    \sqrt{s} = \sqrt{2m_e^2 + 2 E_{-} E_{+} - 2E_{+}k_{-}\cos{\theta_{-}}} 
\end{equation}

\begin{figure}[H]
    \centering
    \includegraphics[width=0.9\linewidth]{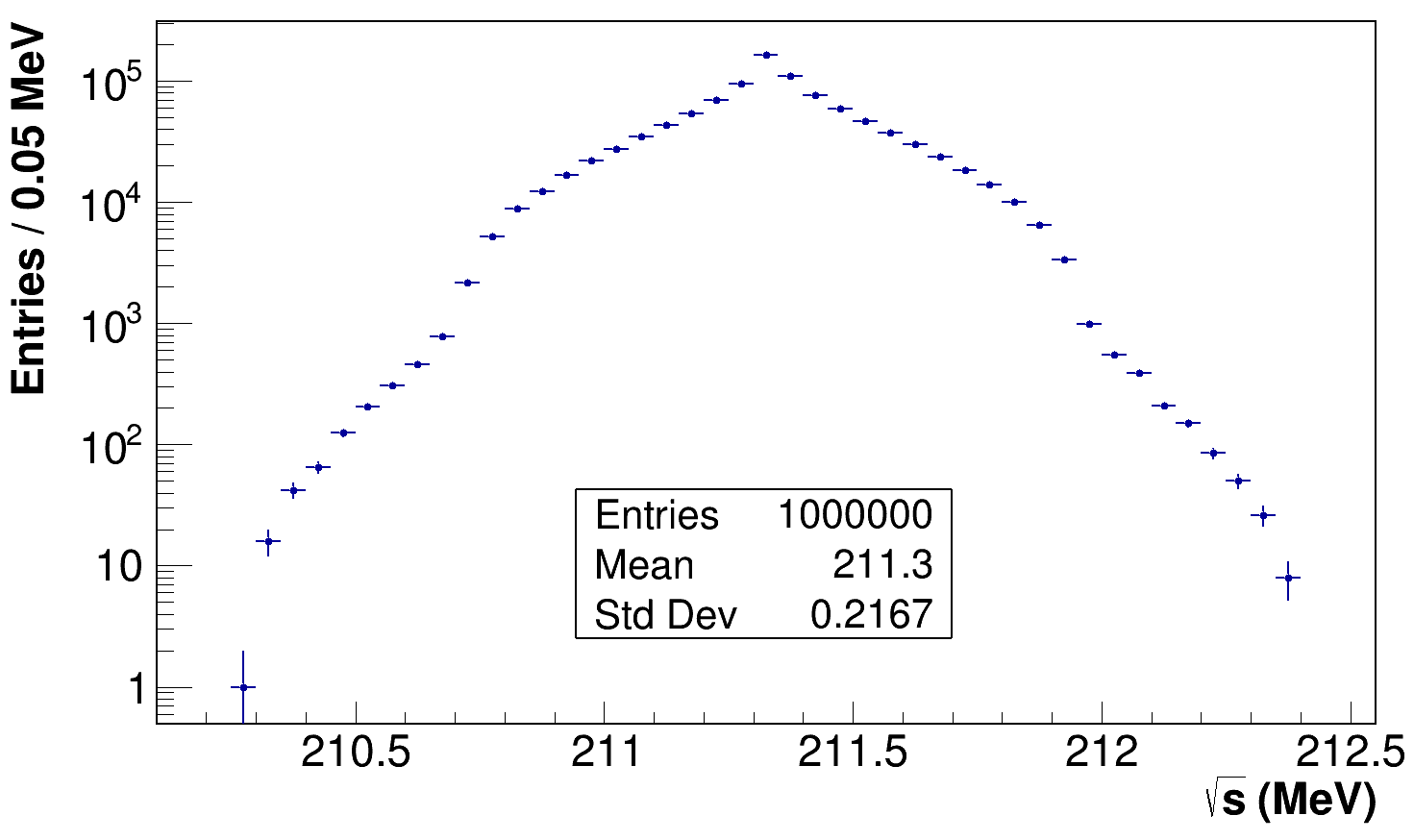}
    \caption{Distribution of $\sqrt{s}$, including only the effect due to target electron motion.}
    \label{fig:nobes}
\end{figure}

The distribution of electron momenta is taken from data obtained with Compton spectroscopy, an experimental technique to prove the momentum-space density $n(k)$ of electrons inside materials, while $\theta$ is distributed isotropically because the lithium targets are polycrystalline.
In the case of lithium metal, the $n(k)$ distribution extrapolated from data is \cite{lithium}:
\begin{eqnarray}
   n(k) =  n_0 - \alpha^- \left(\frac{k}{k_F}\right)^{\beta^-} \quad (k \leq k_F) \\
   = \alpha^+ \left(\frac{k}{k_F}\right)^{\beta^+} \quad (k > k_F) \\
\end{eqnarray}
where $(n_0 = 0.85$), $(\alpha^+, \alpha^-, \beta^+, \beta_-) = (0.144, 0.134, -6.21, 3.24)$ and $k_F = 0.59$ atomic units, corresponding to $k_F c = 2.2$ keV. The resulting $\sqrt{s}$ distribution for a positron beam of $43.7$ GeV, as shown in Figure \ref{fig:nobes}, has a $\sim200$ keV spread, corresponding to a relative fluctuation of 1 $\tcperthousand$.
Moreover, the fluctuation in $\sqrt{s}$ due to the positrons' energy loss in the material is negligible in practical cases: for a total target length of 20 cm, the average energy loss by ionization is $\sim 17$ MeV, translating into  a relative difference in $\sqrt{s}$ which is a factor $O(10^{-1})$ smaller than the fluctuation due to electrons motion.
Hence, in the proposed setup, the dominant effect on the value of the $\sqrt{s}$ of the collisions remains the original 1.2\% beam energy spread typical of the H4 beam line \cite{na64bes}.

\subsection{Beam energy fluctuations}
\begin{figure}[H]
    \centering
    \includegraphics[width=0.9\linewidth]{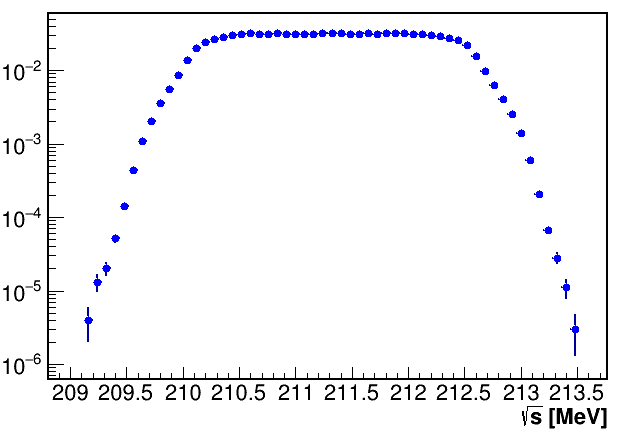}
    \caption{Distribution of $\sqrt{s}$ including lithium target electron motion and beam fluctuations.}
    \label{fig:bes}
\end{figure}

As stated above, the signal yield is reduced by beam energy fluctuations. The beam momentum distribution is assumed to be uniform with a $\pm 1.2\%$ spread \cite{na64bes}, centered at 43.7 GeV.
After including this beam fluctuation, along with the electron motion sketched in the previous section, the full center-of-mass distribution $\mathcal{G}(\sqrt{s})$ was evaluated, as shown in Figure \ref{fig:bes}. As expected, $\mathcal{G}(\sqrt{s})$ resembles the original uniform beam energy distribution, showing signs of the contamination due to electron motion just in the tails.

\subsection{Initial State Radiation}
\label{sectisr}
The combined effects of beam energy fluctuations and ISR (initial state radiation) on TM production cross-section should be carefully evaluated.

The lowest-order cross-section  for producing TM (not including ISR) is assumed to be constant and equal to Eq. \eqref{res} within the window $\left[2 m_\mu - \Delta E, 2 m_\mu \right]$, and zero outside 
this range. Assuming $\mathcal{G}(\sqrt{s})$ the $\sqrt{s}$ probability density function accounting for beam energy spread, and $f_{\text{ISR}}(x; \sqrt{s})$ as the proper QED radiator function (see Appendix \ref{app:isr}), the effective cross-section reads:
\begin{equation}\label{eq:isr-rate}
\scalemath{0.95}{\sigma_{\text{TM,eff.}} = \int d\, s' \,\mathcal{G}_{\text{BES}}(\sqrt{s'}) \int \, dx \, f_{\text{ISR}}(x; \sqrt{s'}) \sigma_{\text{TM}}(x \sqrt{s'})}\,,
\end{equation}
where the $x$ integral is evaluated with the following extrema:
\begin{eqnarray}
x_{\text{min}}(\sqrt{s'}) = \min\left[1, \frac{2 m_\mu - \Delta E}{\sqrt{s'}}\right]
\\
x_{\text{max}}(\sqrt{s'}) = \min\left[1, \frac{2 m_\mu}{\sqrt{s'}}\right]
\end{eqnarray}
By evaluating the integral numerically, a $\sigma_{\text{TM,eff.}}$ of 29 pb, corresponding to a $p$-factor of $4.35 \times 10^{-4}$, is obtained.

\section{Target assembly}
\begin{figure}[H]
    \centering
    \includegraphics[width=0.98\linewidth]{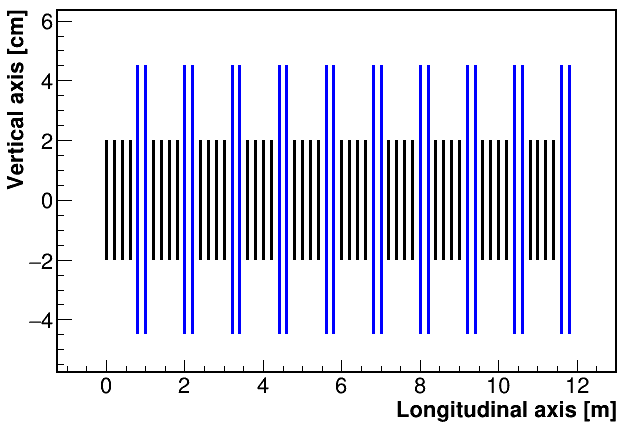}
    \caption{Sketch of the target-tracker setup, including 10 cells with lithium targets in black and silicon detectors in blue.}
    \label{fig:tgt}
\end{figure}

Due to the small value of the cross-section and the relatively low positron fluxes of CERN SPS H4 beamlines, it was decided to study an innovative design with a multiple target assembly, to increase production rates, because with only one target (as proposed in \cite{crivelli}) the signal rate is very low for a discovery, if all effects are included.
The TM dissociation length in lithium is $\mu_{d}^{-1} = 1.86$ mm, which indicates a lithium target thickness of 4 mm ($\sim2\mu_{d}^{-1}$) as the best choice.
The target spacing along the beam direction is designed in such a way that the majority of TM decays occurs between two targets. The number of un-dissociated TM atoms produced for each target per each impinging positron is:
\begin{equation}
        \frac{dTM}{de^+ dN_{\mathrm{target}}} = \epsilon \frac{p \, \sigma_{ON R.}}{13 Z \, \mathrm{b}} (1 - e^{-\Delta z \mu_d}) = 6.6 \times 10^{-13} \epsilon
        \label{ntm_e-13}
\end{equation}
as in Eq. \eqref{ntm}.
As the TM 1S decay length $\beta \gamma c \tau$ for a 43.7 GeV positron beam is 11.3 cm, a reasonable choice, motivated by space constraints in the H4 area, is to have 10 target cells, each featuring 4 targets spaced 20 cm. 
A tracking stations consisting of 2 silicon detectors spaced by 20 cm will be located in between each two cells, for a total of 8 silicon detectors and 40 lithium foils. Including a 20 cm spacing between the last silicon detector of a cell and the first of the next cell, every cell is 120 cm long, for a total length of the target-silicon detectors system of 12 m, as shown in Figure \ref{fig:tgt}.

The target station can fit in the H4A area, before the Goliath magnet, as showed during the LEMMA test beam \cite{tesilemma}. The target's transverse area needs to match the beam spot size (1 cm × 1 cm).

\section{Preliminary backgrounds estimates}
A 43.7 GeV positron beam on a fixed target, can produce both electromagnetic $e^+(e^-,\,p)$ and weak $e^+(e^-,\,p,\,n)$ interactions. 

The rate of electromagnetic Moller $e^+p$ interactions is roughly a factor $m_e/m_p \sim 1/2000$ smaller compared to $e^+e^-$, due to the $1/s$ scaling of cross-sections.

Even when \( e^+p \) interactions occur, the scattered proton typically has low energy because the scattering Møller is dominated by \( t \)- and \( u \)-channel processes. As a result, it rarely produces secondary particles that could potentially mimic a signal.

The Bhabha scattering cross-section is tens to hundreds of microbarn, depending on the angular cuts \cite{gargiulo2024true}, while the weak cross-sections are of the order of 1 pb per atom \cite{pdg}.


Bhabha scattering $e^+e^- \to e^+e^-$ is the main source of background, as indicated by the comparison of cross-sections. Additionally, it is the only background source that shares the same center-of-mass energy and final states as the target process (TM)  $e^+e^-$ decays. Experimentally, the primary differences between Bhabha scattering and TM decays are the angular distributions in the center-of-mass frame and the displaced decays of TM.

Therefore, the background suppression is divided in three steps: selecting Bhabha+TM events, suppressing the Bhabha background, by applying angular cuts, and finally isolating TM decays, by leveraging their displaced decay vertices.

\subsection{Bhabha scattering background estimate}

\begin{figure}[H]
\centering
    \includegraphics[width=\columnwidth]{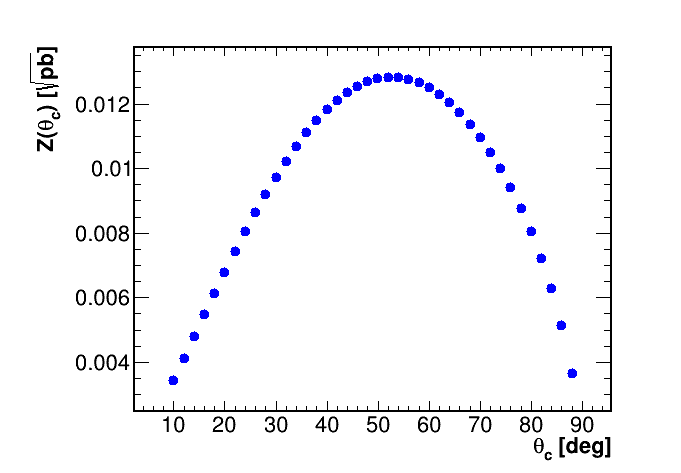}
    \caption{Significance scan in the cut angle $\theta_c$ ($\theta_{cm} \in [\theta_c,\,\pi - \theta_c]$), at $1^\circ$ steps (see Eq. \eqref{signtheta}). The peak is around $53^\circ$.}
    \label{fig:theta}
\end{figure}

At the leading order, the differential Bhabha scattering cross-section is given by:
\begin{small} 
\begin{equation}
\label{bhabhaeq}
    \dfrac{d\sigma}{d\Omega^*} = \dfrac{\alpha^2}{2s} \left[\dfrac{1 + \cos^4(\theta_{cm}/2)}{\sin^4(\theta_{cm}/2)} - \dfrac{2\cos^4(\theta_{cm}/2)}{\sin^2(\theta_{cm}/2)} + \dfrac{1 + \cos^2(\theta_{cm})}{2}\right]\end{equation}
    \end{small} 
Electron pairs originating from Bhabha scattering have predominantly small $\theta_{cm}$ angles. 
TM decay products, on the contrary, are distributed as expected for spin-1 particles: 
$\dfrac{dN}{d\cos{\theta_{cm}}} \propto \left( 1 + \cos^2{\theta_{cm}} \right)$ .
For simplicity, a symmetrical angular cut $\theta_{cm} \in [\theta_c,\,\pi - \theta_c]$ was chosen to partly discriminate signal from background. 
The asymptotic significance, shown in Figure \ref{fig:theta}, is \cite{pratstat}:
\begin{equation}
    Z(\theta_c) = \frac{\sigma_{\mathrm{TM}}(\theta_c < \theta_{cm} < \pi - \theta_c)}{\sqrt{\sigma_{\mathrm{Bhabha}}(\theta_c < \theta_{cm} < \pi - \theta_c)}},
    \label{signtheta}
\end{equation}
 The significance was scanned for several $\theta_c$ values. The shape of $Z(\theta_c)$ does not change if the signal or background yields are modified by other uncorrelated quantities. For this reason, its maximum could be used to identify the optimal angular cut. 
 As a compromise between the optimal angle $\theta^c_{\text{opt}} = 53^\circ$ and the desire to maximize the TM yield, an angular cut of $\theta_c = 45^{\circ}$ was chosen. This decision resulted in a reduction of the signal yield by a factor of $\epsilon_{\theta_{cm}}$ = 62\% and a Bhabha scattering cross-section of $\sigma_{Bh.} = 21 \mu b $.
 The minimum (maximum) angles in the lab frame of the $e^+/e^-$ originating from TM decays or Bhabha scattering are then 2.7(16.6) mrad, corresponding to maximum (minimum) energies of 37.3(6.4) GeV.

\section{Detector requirements}

\begin{figure}[H]
    \centering
    \includegraphics[width=1\linewidth]{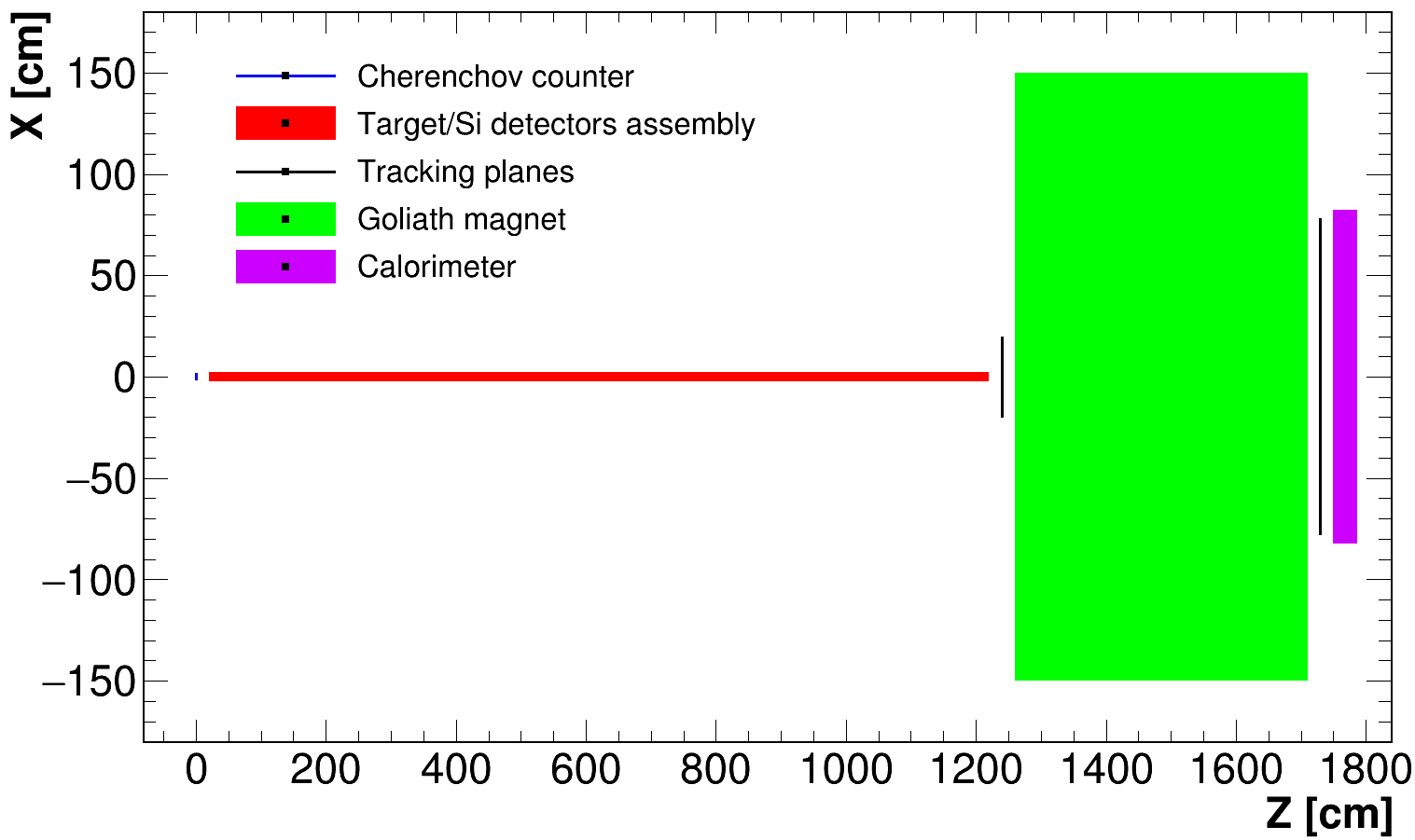}
    \caption{Sketch of the setup including the detectors and the Goliath magnet.}
    \label{fig:disegno}
\end{figure}

In order to efficiently select TM events, the momentum and decay vertex of its decay products have to be measured.
To complete this task, a minimal set of detectors is needed, as shown in Figure \ref{fig:disegno}:
\begin{itemize}
    \item A gas Cherenchov threshold counter before the target, to distinguish beam positrons from hadron contamination, with a purity to be established with further simulations
    \item A target assembly equipped with silicon pixel based vertex detectors
    \item A large-area spectrometer, with tracking planes before (just at the end of target) and after the Goliath magnet at H4A \cite{goliath} for charge selection and photon rejection
    \item An electromagnetic calorimeter downstream of the spectrometer
\end{itemize}

\subsection{Silicon trackers within the target assembly}
The silicon detectors area must be dimensioned on the basis of the beam spot at H4 ($\sim 1\times1$ cm$^2$) and of the maximum transverse track projection after 1 cell (1 m), i.e., $16.6$ mrad $\times$ 1 m = 1.66 cm. 
According to these values, silicon detectors with an area of $\sim 4.5\times4.5$ cm$^2$ provide full geometrical acceptance. \\
As indicated by simulations, a per-layer 5 $\mu$m resolution and a 0.3\% $X_0$ material budget are required. Only very thin monolithic pixel sensors, such as the ones foreseen for the ITS-3 ALICE upgrade \cite{its3} are able to match requirements. The total silicon detectors area would be 405 cm$^2$ that is $\sim$0.04 m$^2$, to be compared to the $\sim$10 m$^2$ total area of ITS-3, whose cost is estimated around $\sim 5$ million CHF \cite{its3price}. Scaling the price naively with the detector's area, a relatively contained cost around 20 kCHF can be estimated.

\subsection{Spectrometer and calorimeter}
H4 beam line has some permanent equipment among which a large gap dipole magnet known as Goliath which is frequently used as the magnetic element of spectrometers.
The Goliath magnet features a uniform vertical magnetic field of  B = 1.2  T over a length of 2 m along the beam axis, with an aperture of 2.3 m in the  x -direction and 0.9 m in the  y -direction \cite{goliath}.


With such high bending power, even the most energetic particles have a radius of curvature of $R \sim 120$ m. The resulting horizontal angle of $L/R = 16.4$ mrad, translates into a difference $\Delta x_{min}=7.4$ cm between $x$-coordinates of the tracks before the 4.5 m long magnet and expected positions in $x$ after the magnet.
By using two large-area scintillating or gas detectors before (just at the end of the target) and after the magnet, positive and negative particles can be easily discriminated.\\

With a $x$ coordinate resolution $\sigma_x = 0.5$ cm, negative and positive particles can be separated with at least $2 \Delta x_{min}/(\sqrt{2}\sigma_x) = 20\sigma$ confidence.\\

By using a low material-budget detector for the first tracking plane to prevent photon conversions, this setup can effectively reject photons. If photons convert in the first tracker, they produce an additional electron-positron pair which is vetoed, while if they convert in the second tracking plane, they produce hits not matching the first tracker hits. If no conversion occurs in the two planes, photons can be easily rejected by matching hits between the calorimeter and tracking detectors, as it is typically done.\\

A key aspect of the feasibility study is the impact of the magnet on the angular acceptance.
After the angular cuts chosen above, the maximum transverse projection of tracks at the end of the target is $16.6$ mrad $\times 12$ m  $\sim 20$ cm, fitting the magnet aperture in $x$ and $y$.
On the other hand, after the magnet (placed 16.5 meters after the first target), the maximum projection is 27.4 cm neglecting the magnetic field effect, which provides an additional spread in $x$ of 7.4 cm for 43.7 GeV tracks and 50.5 cm for tracks at the minimum energy of 6.4 GeV, therefore fitting in any case the magnet aperture in $x$ and $y$.\\

In order to achieve 100\% geometrical acceptance, the two tracking planes must have active areas about $\sim$40 cm $\times$ 40 cm and 55 cm $\times$ 156 cm, respectively. 
The calorimeter should cover the same area as the second detector. A cost-effective solution for the calorimeter is to use large lead-glass blocks, similar to those used in the OPAL experiment, which can provide an energy resolution of  $\frac{5\%}{\sqrt{E\,[\text{GeV}]}}$   \cite{opal}.

\section{Monte Carlo simulations}

A proof-of-concept simulation was performed to demonstrate the possibility to efficiently suppress the background, by distinguishing TM decays with displaced vertices from events (mostly Bhabha) originating from the targets.
The simulation used the Geant4 package and employed only four $15\times15$ mm$^2$ targets and silicon detectors (with a 300 $\mu$m thickness each) for one cell, and a $20\times20$ cm$^2$ virtual detector at 20 cm from the last silicon array, to mimic the spectrometer-calorimeter setup.
A total of $N_{POT} = 10^{14}$ positrons on target (POT) were simulated, using as a monochromatic pencil beam with no beam spot size. \\

Due to computing power constraints, a full simulation with all 10 cells was not feasible. Also a simulation including all the detectors are outside the scope of this paper. 
However, Bhabha scattering and TM decay products only cross a few cells for geometrical reasons, making the one-cell simulation an acceptable first estimate. 
Each cell has a $\sim$1\% $X_0$ material budget, therefore the effect on the signal efficiency is negligible. An increase of the background is also unlikely, given that interactions of Bhabha scattering products in cells after detection cannot spoil in any way the vertex already reconstructed by the silicon trackers, 
considering in addition that the analysis cuts are designed to reject hard interactions that are not 2-body processes.\\

A pre-selection was applied during the simulation based on the number of tracks impinging on the virtual detector.
Only events with exactly 1 positive and 1 negative charged tracks, and any number of neutral particles on the detector were accepted. In addition, both charged particles were required to have 2 mrad$<\theta_{lab}<$20 mrad and energies between 3 and 42 GeV. 
After generation stage, to simulate the experimental effects, a smearing, reconstruction and selection procedure was applied to saved data. 

Particle energies were smeared with a realistic calorimeter resolution of $\sigma_E/E = 5\%/\sqrt{E[\mathrm{GeV}]} \oplus 10\%/E[\mathrm{GeV}] \oplus 1\%$.

The $x$ and $y$ positions on silicon detectors were smeared with a Gaussian 5 $\mu$m resolution, and finally $\theta_{lab}$ angles were reconstructed using information from the smeared positions on the silicon detectors. 

To identify Bhabha (TM-like) events within the $\theta_{cm}$ acceptance, a pre-selection was first applied, requiring:
\begin{itemize}
\item Total energy within 15\% of the beam energy
\item Total energy of charged particles within 30\% of the beam energy
\item Two tracks in each silicon detector
\item Exactly one positive and one negative track in the virtual detector after the target
\item $\theta_{lab} > 2$ mrad for both tracks
\item Combined mass of the two tracks within 15 MeV of the TM mass
\end{itemize}

The cumulative efficiency of these cuts is estimated to be 92\%, by normalizing with respect to the total number of Bhabha events expected with 4 targets in the $\pi/4 < \theta_{cm} < 3/4 \pi$ acceptance, i.e.:
\begin{equation}
    N_{Bh.} = N_{target} N_A \rho \frac{Z}{A} \Delta z \, \sigma_{Bh.} N_{POT} = 4.66 \times 10^{-6} N_{POT}
\end{equation}

The separation of positive, negative and neutral particles is obtained using the combined spectrometer-calorimeter geometry as already described, but the impact of photons converting in the tracking planes should be evaluated with further simulations.

The $z$ position of the starting point of each track was reconstructed as $\sqrt{x^2+y^2}/\arctan{\theta}$, where $x,y$ and $\theta$ are measured by the silicon pixel trackers. The $z$ coordinate of the vertex is then evaluated as the arithmetic average of $z^+$ and $z^-$, i.e., the $e^+$ and $e^-$ reconstructed $z$ positions. It was checked that this simple and faster way of reconstructing $z$ gives identical results for our purposes with respect to a two-tracks fit procedure requiring a common vertex.

After $z$-reconstruction, additional quality cuts were applied, as shown in Table \ref{tab:eff}. The resulting total selection efficiency $\epsilon_{reco}$ is $77.4\%$.
\begin{table}[H]
\setlength{\tabcolsep}{4pt}
\begin{tabular}{lccc}
Cut                                                  & $\epsilon$ & $\frac{\#Selected}{\#Bhabha}$ & $\frac{\#Selected}{\#Pre-sel.} $
\\\hline
Pre-selection                                        & N.A.           & 91.8\%                         & 100.0\%                              \\
$|z^+ - z^-|$ \textless 4.5 cm                    & 98.8\%         & 90.6\%                          & 98.8\%                                \\
$|p_x^+ + p_x^-| < 8$ MeV                                       & 94.4\%         & 85.5\%                          & 93.2\%                                \\
$|p_y^+ + p_y^-| < 8$ MeV                                        & 95.9\%         & 82.0\%                          & 89.3\%                                \\
$|E^+ + E^-| - E_{\mathrm{beam}}$ \textless \, 2 GeV & 97.5\%         & 79.9\%                          & 87.1\%                                \\
$|\theta_{cm} - \frac{\pi}{2}| < \frac{\pi}{4}$                  & 96.9\%         & 77.4\%                          & 84.4\%                                               
\end{tabular}
\caption{Summary of the cuts applied on simulated events, and their efficiency $\epsilon$ with respect to the previous one, to the number of expected Bhabha events, and pre-selected events.}\label{tab:eff}
\end{table}

\begin{figure}[H]
    \centering
    \includegraphics[width=\linewidth]{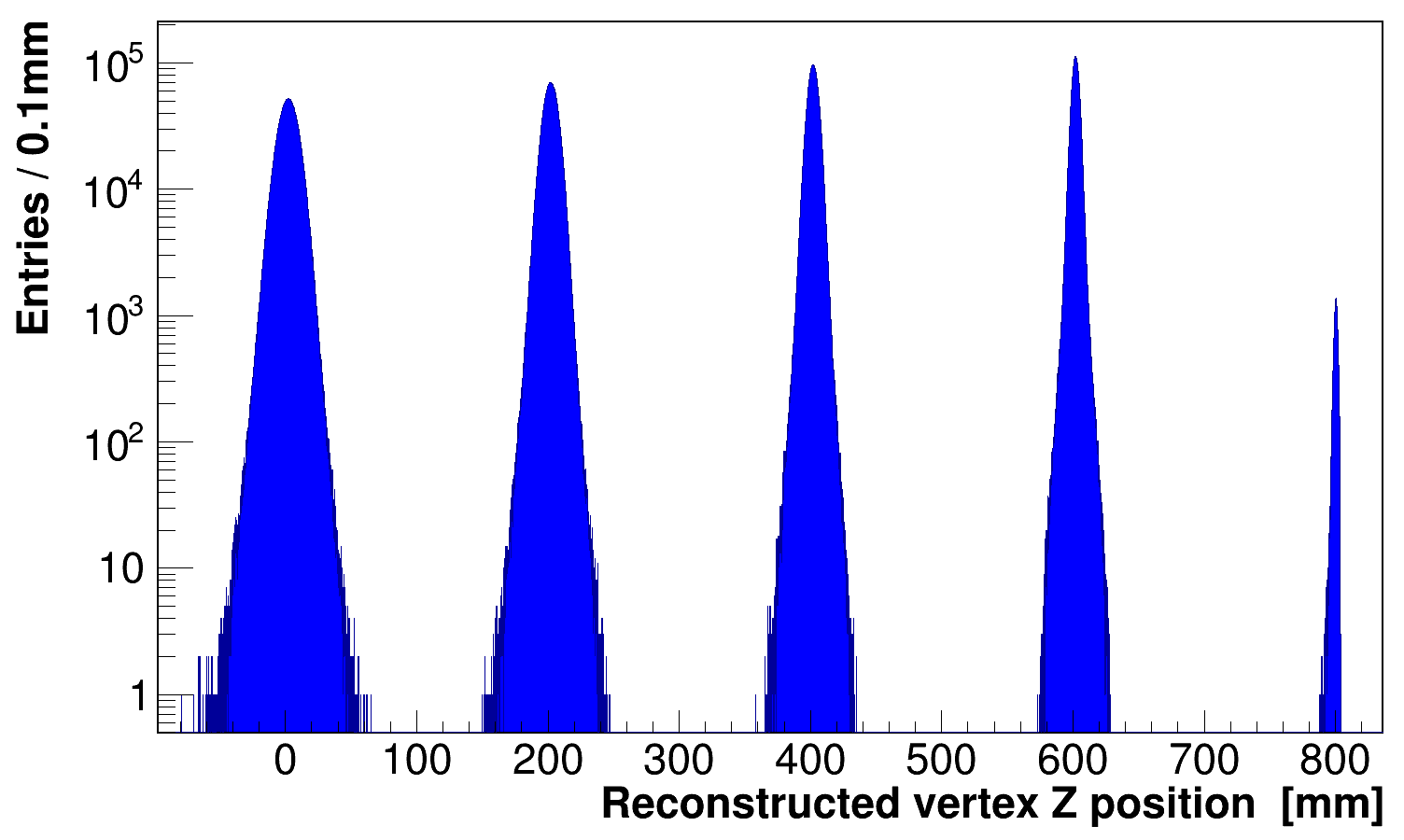}
    \caption{Reconstructed vertex $z$ position  after all other cuts for 4 targets of one simulated cell. The data at 800 mm provide a very small yield and are due to fake vertices inside the first silicon detector.}
    \label{fig:z}
\end{figure}

The TM-Bhabha scattering separation is obtained using cuts on the reconstructed $z$ coordinated of the vertex (see Fig. \ref{fig:z}). Only the regions in $z$ with no background in the simulation are accepted, resulting in an expected background rate of  $10^{-14}$ events per POT, based on the simulated statistics.

The efficiency of the cuts on the vertex for each target (see Table \ref{tab:zcuts}), summed for the 4 targets on the simulated cell, is 1.7, therefore the average efficiency of the vertex-based selection is $\epsilon_{v} = 1.7/4 = 42.5\%$. 

\begin{table}[H]
\centering
\begin{tabular}{cccc}
Target $z$ {[}mm{]} & $z_{min}$ {[}mm{]} & $z_{max}$ {[}mm{]} & Partial $\epsilon_v$ \\
\hline 0                        & 70                   & 150                  & 27.3\%               \\
200                      & 250                  & 356                  & 39.1\%                \\
400                      & 438                  & 571                  & 49.4\%               \\
600                      & 631                  & 782                  & 56.0\%              
\end{tabular}
\caption{Cuts ($z_{min}, \, z_{max}$) on vertex $z$ position for each target of one simulated cell, with the corresponding efficiencies, evaluated as the integral of the exponential probability for TM decay.}\label{tab:zcuts}
\end{table}

\section{Discovery potential}
The global efficiency in equation \eqref{ntm_e-13} is evaluated by combining the probability to produce a 1S TM ($\epsilon_{1S} = 83\%$), the angular efficiency ($\epsilon_{\theta_{cm}} = 62\%$), the reconstruction efficiency ($\epsilon_{reco} = 77.4\%$), and the vertex-based selection efficiency ($\epsilon_v = 42\%$), reaching a value of $\epsilon = 16.5\%$. After multiplying by the number of targets ($N_{target} = 40$), the value of selected TM per POT is evaluated as:
\begin{equation}
\frac{\#TM}{\#e^+} = \epsilon N_{target} \cdot 6.6 \times 10^{-13} = 4.35 \times 10^{-12}
\end{equation}
The expected background yield per POT depends on the simulated statistics ($10^{14}$) for one cell and the number of cells ($N_{cells}=10$):
\begin{equation}
 \frac{\#BKG}{\#e^+} = N_{cells} \cdot 10^{-14}= 10^{-13}
\end{equation}
For H4, about 3000 spills/day are expected \cite{dune}, with 
with a 4.8 s duration and a maximum intensity of $10^7$ positrons per spill at 100 GeV \cite{na642023}. A test beam for the LEMMA muon production scheme was performed at $5\times 10^6$ $e^+$ per spill, without exploiting the maximum intensity \cite{tesilemma}, at an energy close to the required value for TM production (43.6 GeV), while the NA64 collaboration quotes rates between $5\times 10^6$ $e^+$ and $7\times 10^6$ $e^+$ per spill at 100 GeV. Given that the positron beam production efficiency increases at lower energies \cite{sps}, a rate of $5\times 10^6$ $e^+$ per spill is taken as a conservative value, and a rate of $10^7$ $e^+$ per spill is taken as an optimistic one.

Therefore, in 3 months of data-taking, a total of $2.7$ $(5.4) \times 10^5$ spills, corresponding to $\#e^+ = 1.35$ $(2.7) \times 10^{12}$, are integrated with conservative (optimistic) assumptions on the positron rate.
This translates in $5.8$ ($11.6$) expected signal events and 0.13 (0.26) background events with conservative (optimistic) assumptions, corresponding to a significance of 5.8 (8.2) $\sigma$.
Note finally that the expected background is likely overestimated, due to the relatively low statistics of the simulated sample.

\section{Conclusion}
Among QED-bound states, one of the most interesting ones is the so-called true muonium (TM), a $\mu^+\mu^-$ bound state, never observed so far. Ortho-TM (its spin-1 state) can be produced from $e^+ e^-$ interactions on resonance with a 67 nb peak cross-section at a fixed target experiment employing a 43.7 GeV positron beam.
Taking into account the beam energy spread, the very small width of the TM resonance, a O(1\%) beam energy spread, and initial state radiation effects, the cross-section is reduced to $O(10)$ pb.

In this paper, we explored the possibility of searching for TM in positron on target collisions at the CERN North-Area H4 beam line using a multi target approach. Each of the 10 target station includes 4 lithium targets followed by two very thin silicon detectors. According to the preliminary calculations and simulations described in this paper, a discovery can be obtained in few months of data taking.


\section{Acknowledgements}
The authors are grateful to M. Raggi for the careful reading of the paper and for valuable suggestions and comments.

\appendix
    \section{Methods}
\label{methods}
The significance is calculated as: 
\begin{equation}
    Z = \sqrt{-2 \log{L(N, 0)/L(N, 1)}}
\end{equation}
where $L(N, \mu)$ is the Poissonian likelihood with $N$ observed events, a signal strength of $\mu$ (0 for background only, 1 for nominal signal yield), $s$($b$) expected signal (background) events \cite{pratstat}
\begin{equation}
    L(n, \mu) = \frac{(\mu s + b)^N}{N!} \exp{-(\mu s + b)}
\end{equation}
    Therefore $Z$ can be rewritten in the simpler formula:
\begin{equation}
 Z = \sqrt{2 \left[ (s+b) \log \left( 1 + \frac{s}{b}\right)   -s \right]}    
\end{equation}

\section{Initial State Radiation} \label{app:isr}
The ISR radiator function used in Eq. \eqref{eq:isr-rate} is essentially the probability that the electron pair carries a given fraction of the nominal center of mass energy. The following relations were used: \cite{greco, jadach1, jadach2}
\begin{equation}
f_{\text{ISR}}(x;s) = f^0_{\text{ISR}}(x; s)\,\left(1 + \frac{\beta_l}{2} - \frac{1}{2}(1-x^2)\right)\,,
\end{equation}
where $\beta_l = \frac{2\alpha}{\pi}\left(\log{\frac{s}{m_e^2}} - 1 \right)$, and
\begin{equation}
f^0_{\text{ISR}}(x;s) = \frac{\exp{\left(\frac{\beta_l}{4} + \frac{\alpha}{\pi}\left(\frac{1}{2} + \frac{\pi^2}{3}\right)-\gamma_E \beta_l\right)}}{\Gamma(1+\beta_l)} \beta_l(1-x)^{\beta_l - 1}\,.
\end{equation}

\bibliography{apssamp}

\end{document}